\documentstyle[twoside,fleqn,espcrc2my]{article}

\newcommand{\nn} {\nonumber}
\newcommand{\nc}{\newcommand}
\newcommand{\rnc}{\renewcommand}

\def\={\ =\ }
\def\+{\ +\ }
\def\-{\ -\ }

\nc{\Tr}{{\rm Tr\,}}
\nc{\rome}{{\rm Roma}}
\nc{\ie}{{\em i.e.}}
\nc{\eg}{{\em e.g.}}
\nc{\etal}{{\em et al.}}
\nc{\calF}{{\cal F}}
\nc{\calL}{{\cal L}}
\nc{\calS}{{\cal S}}
\nc{\calO}{{\cal O}}
\nc{\kapbar}{\bar{\kappa}}
\nc{\uidot}{\dot u}
\nc{\uiidot}{\ddot u}
\nc{\uiiidot}{\stackrel{\ldots}{u}}
\nc{\psii}{\psi^{(1)}}
\nc{\psibar}{\overline\psi}
\nc{\chibar}{\overline\chi}
\nc{\psiibar}{\overline\psi^{(1)}}
\nc{\psibari}{\overline\psi^{(1)}}

\rnc{\topfraction}{1.0}
\rnc{\bottomfraction}{1.0}
\rnc{\textfraction}{0.0}

\nc{\qq}{\P}
\nc{\qqc}{[\P]}

\nc{\rng}{\rangle}
\nc{\lng}{\langle}
\nc{\rcite}{ref.\ \cite}
\nc{\ba}{\begin{array}}
\nc{\ea}{\end{array}}
\nc{\lb}{\left(}
\nc{\rb}{\right)}
\nc{\qrt}{\frac{1}{4}}

\nc{\al}{\alpha}
\nc{\bt}{\beta}
\nc{\gm}{\gamma}
\nc{\dl}{\delta}
\nc{\ep}{\epsilon}
\nc{\varep}{\varepsilon}
\nc{\zt}{\zeta}
\nc{\et}{\eta}
\nc{\th}{\theta}
\nc{\kp}{\kappa}
\nc{\lm}{\lambda}
\nc{\rh}{\rho}
\nc{\sg}{\sigma}
\nc{\ta}{\tau}
\nc{\ph}{\phi}
\nc{\vr}{\varphi}
\nc{\ch}{\chi}
\nc{\ps}{\psi}
\nc{\om}{\omega}
\nc{\noi}{\noindent}
\nc{\half}{\frac{1}{2}}
\nc{\rr}[1]{$^{#1}$}
\nc{\rf}[1]{(\ref{#1})}
\nc{\rfs}[2]{(\ref{#1},\ref{#2})}
\nc{\smgr}{\stackrel{\textstyle <}{>}}
\nc{\grsm}{\stackrel{\textstyle >}{<}}
\nc{\aleq}{\mbox{}_{\textstyle \sim}^{\textstyle < }}
\nc{\ageq}{\mbox{}_{\textstyle \sim}^{\textstyle > }}
\nc{\ra}{\rightarrow}
\nc{\lra}{\leftrightarrow}
\nc{\be}{\begin{equation}}
\nc{\ee}{\end{equation}}
\nc{\bea}{\begin{eqnarray}}
\nc{\eea}{\end{eqnarray}}
\nc{\eqrf}{eq.\ \rf}
\nc{\erf}{{\rm erf}}

\nc{\ap}[1]{Ann.\ Phys.~#1\ }
\nc{\app}[1]{Acta Physica Polonica~#1\ }
\nc{\arnps}[1]{Ann.\ Rev.\ Nucl.\ Part.\ Sci.~#1\ }
\nc{\cmp}[1]{Commun.\ Math.\ Phys.~#1\ }
\nc{\cpc}[1]{Comput.\ Phys.\ Commun.~#1\ }
\nc{\jetp}[1]{Soviet Physics JETP~#1\ }
\nc{\jpa}[1]{J.\ Phys.\ A~#1\ } 
\nc{\jpg}[1]{J.\ Phys.\ G~#1\ } 
\nc{\mpla}[1]{Mod.\ Phys.\ Lett.~A#1\ }
\nc{\npa}[1]{Nucl.\ Phys.~A#1\ }
\nc{\npb}[1]{Nucl.\ Phys.~B#1\ }
\nc{\nproc}[1]{Nucl.\ Phys.~B (Proc.\ Suppl.)~#1\ }
\nc{\pla}[1]{Phys.\ Lett.~#1A\ }
\nc{\plb}[1]{Phys.\ Lett.~#1B\ }
\nc{\pr}[1]{Phys.\ Rep.~#1\ }
\nc{\pra}[1]{Phys.\ Rev.~A#1\ }
\nc{\prb}[1]{Phys.\ Rev.~B#1\ }
\nc{\prc}[1]{Phys.\ Rev.~C#1\ }
\nc{\prd}[1]{Phys.\ Rev.~D#1\ }
\nc{\pre}[1]{Phys.\ Rev.~E#1\ }
\nc{\prep}[1]{Phys.\ Rep.~#1\ }
\nc{\prev}[1]{Phys.\ Rev.~#1\ }
\nc{\prl}[1]{Phys.\ Rev.\ Lett.~#1\ }
\nc{\procroy}[1]{Proc.\ Roy.\ Soc.~#1\ }
\nc{\ptp}[1]{Prog.\ Theor.\ Phys.~#1\ }
\nc{\rmp}[1]{Rev.\ Mod.\ Phys.~#1\ }
\nc{\rpp}[1]{Rep.\ Prog.\ Phys.~#1\ }
\nc{\sjnp}[1]{Sov.\ J.\ Nucl.\ Phys.~#1\ }

\newcommand{\AmS}{{\protect\the\textfont2
  A\kern-.1667em\lower.5ex\hbox{M}\kern-.125emS}}

\title{Laplacian Abelian Projection:\\ Abelian dominance and Monopole
       dominance\thanks{Contribution to Lattice '98, Boulder, CO, USA, July '98.\protect\\
                        ETH-Z\"urich preprint {\bf SCSC-TR-98-08}.}}

\author{A.J.\ van der Sijs\address{Swiss Center for Scientific Computing,
        ETH-Z\"urich, ETH-Zentrum, CH-8092 Z\"urich,
        Switzerland\\ ({\tt arjan@scsc.ethz.ch})}
       }

\begin{document}

\input epsf
\epsfverbosetrue

\begin{abstract}
A comparative study of Abelian and Monopole dominance in the Laplacian
and Maximally Abelian projected gauges is carried out.
Clear evidence for both types of dominance is obtained for the Laplacian
projection. Surprisingly, the evidence is much more ambiguous in
the Maximally Abelian gauge. This is attributed to possible
``long-distance imperfections'' in the maximally abelian gauge fixing.
\end{abstract}

\maketitle

\section{INTRODUCTION}

Despite its many successes, the Maximally Abelian Gauge (MAG) \cite{mag}
has the great drawback that it is {\em ambiguous}.
A precise way to phrase this ambiguity is as follows:
it is in general unlikely, and certainly impossible to guarantee,
that the configuration obtained by the
usual local iterative minimization algorithm be (arbitrarily) close to
the desired configuration
$\{ \bar U_{\mu,x} \! = \bar\Omega_x U_{\mu,x}
\bar\Omega^+_{x+\hat\mu}\}$,
no matter how high the numerical precision of the computer and no
matter how long the iteration is continued.
(The reason is well known: one may get stuck in a local minimum.)\@
Here $\{ \bar\Omega_x \}$ is the unknown, true (absolute) minimum of
the functional
\be
\tilde\calS_U(\Omega) \= \sum_{x,\mu} \left\{ 1 -
 \half \Tr \left[\sigma_3 U_{\mu,x}^{(\Omega)}
            \sigma_3 U_{\mu,x}^{(\Omega)\,+} \right] \right\} .
\label{S1}
\ee

As a consequence, a different result is obtained if the
procedure is applied to the same configuration several times,
starting from a different (random) gauge each time \cite{born,suzukiambig}.
In physical terms: {\em gauge covariance is lost}.

The Laplacian Abelian Gauge (LAG) \cite{lat96,ykis97} solves this problem.
This is a unique and unambiguous gauge fixing prescription which can
be pursued to arbitrarily high precision.
Gauge covariance is guaranteed, and one has control
over numerical errors.

In addition, LAG leads to at least as smooth configurations as MAG.
This is important for a reliable extraction of Abelian continuum gauge fields
$A_\mu(x)$.
In fact, in Ref.~\cite{ykis97} it was argued that the fields in the LAG
can be considered to be smoother than in the MAG, as physical monopoles
are treated more ``respectfully'' in the Laplacian gauge.

The present contribution focuses on abelian and monopole dominance
\cite{suzyot,shibasuzuki} in LAG and MAG.

\section{THE LAPLACIAN METHOD}

The minimization problem of Eq.~\rf{S1} can be viewed as the minimization
of the gauge covariant kinetic energy of a real adjoint scalar field
$\phi^a$ ($a=1,2,3$).
In continuum notation \cite{thesis,lat96,ykis97}:
\be
\tilde\calS_A(\phi) \ = \
\int_V \half (D_\mu \phi)^2  \, .
 \label{glass}
\ee
The ambiguities in the MAG arise because of the constraints $|\phi(x)|=
\sum_{a=1}^3 (\phi^a)^2 = 1$.

The idea of the Laplacian gauge fixing is to relax the latter constraint.
Minimization of (\ref{S1},\ref{glass}) then amounts to determining the
lowest mode of the covariant Laplacian $-(D_\mu)^2$.
The corresponding lowest eigenvector $\phi_0$ determines the gauge
transformation to be applied to the gauge field configuration.
Subsequently, the abelian projected fields can be extracted in the
standard way.

The computation of the lowest eigenmode can be done to arbitrary precision,
using standard sparse matrix routines (Lanczos, Rayleigh-Ritz).
The only ambiguity arises when the lowest eigenvalue is degenerate.
This would signal a {\em true\/} Gribov copy.
In practice, however, this never occurs.

For further details, see Refs.\ \cite{lat96,ykis97}.

\section{ABELIAN DOMINANCE AND\protect\\ MONOPOLE DOMINANCE}

We consider a set of 18 pure SU(2) configurations on a $16^4$ lattice
at $\beta = 2.5$.
These configurations are gauge fixed using MAG and LAG, and for both
gauge fixed configurations the abelian
field components are extracted, and the elementary-cube monopoles are
identified using the standard prescription.
In this way we can compare results in the two gauges on the same set
of SU(2) configurations.

Wilson loops and Creutz ratios are calculated from the original (``full'')
non-abelian SU(2) configurations, from the abelian fields in both gauges,
and from the abelian fields generated by the monopole content only.
The Creutz ratios are defined as
\bea
\lefteqn{\chi(R+\half,T+\half) \ =\ } \nn \\
  && - \ln \frac{\langle W(R,T)     \rangle \,
              \langle W(R+1,T+1) \rangle}
             {\langle W(R,T+1)   \rangle \,
              \langle W(R+1,T)   \rangle}
 \, ,
 \label{creutz}
\eea
where $W(R,T)$ denotes an $R\times T$ Wilson loop.
Statistical errors have been determined by means of a bootstrap analysis
on each of the Creutz ratios $\chi(R+\half,T+\half)$ separately.

The MAG fixing was done using a standard iterative algorithm, with alternating
cooling and overrelaxation sweeps, and a very tight stopping criterion:
the iteration was terminated when $\half\,$Tr of the gauge transformation
matrix connecting subsequent configurations deviated from unity less than 
$10^{-12}$ {\em at each site}.
\begin{figure}[bt]
\centerline{
\epsfxsize=\columnwidth
\epsfbox{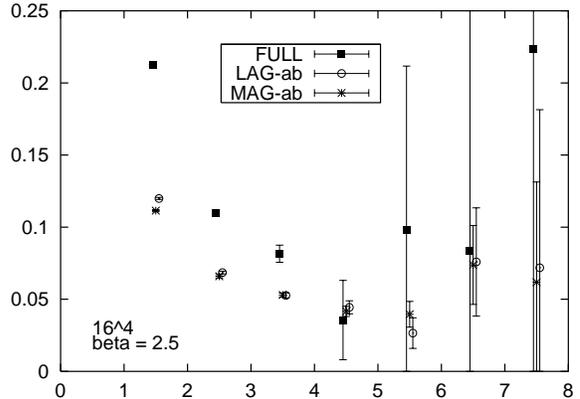}
}
\vspace*{-5mm}
\caption{
Abelian dominance:
Diagonal Creutz ratios $\chi(L+\half,L+\half)$ against $L+\half$.
Shown are full SU(2) and abelian Creutz ratios in MAG and LAG.
}
\vspace*{-2mm}
\label{fig1}
\end{figure}
 
{\em Abelian dominance.}
Fig.~1 shows diagonal Creutz ratios for the full and the abelian projected
theories.
Although the large-distance data are fairly noisy, the trend is the same as
in the work of other authors:
the full and abelian data sets seem to approach a single plateau at large
distances, in agreement with a linearly confining potential, and with abelian
dominance.

It is interesting to compare the two abelian projections,
MAG and LAG.
The data sets almost coincide, but there is a small but significant
difference, at least in the short-distance regime: the LAG data lie
closer to the full SU(2) data than the MAG data, suggesting quantitatively
stronger abelian dominance when the Laplacian Abelian Projection is used.
It would be very interesting to see how this behaviour evolves at
larger distances.

{\em Monopole dominance.}
Before presenting the results, let us discuss the important point that
the monopole gauge field cannot be computed
from the usual monopole currents $k_\mu$ alone.
(See Ref.~\cite{smitsijs}, Eqs.\ (24), (30).)\@
In Landau gauge one has
\be
A^{\rm mon}_\mu(x) = -2\pi\sum_y D(x-y)\,
   \partial_\nu'{\tilde m}_{\nu\mu}(y) ,
\label{Amon}
\ee
where ${\tilde m}_{\nu\mu}$ is the dual of the ``Dirac sheet'' field
$m_{\nu\mu}$, and $D(x-y)$ is the 4-dimensional Coulomb propagator.
It is this gauge field $A^{\rm mon}_\mu$ which determines the monopole
Wilson loop.
In addition to the contribution from the monopole currents $k_\mu(x)$,
given by
\be
k_\mu(x) \ = \ \partial_\nu m_{\nu\mu}(x) ,
\label{kmu}
\ee
there is an additional contribution from the zero mode in $m_{\nu\mu}$.
If this zero-mode contribution is ignored, a `trivial' $N_s\,\times\,N_t$
Wilson loop will not equal unity.
\begin{figure}[bt]
\centerline{
\epsfxsize=\columnwidth
\epsfbox{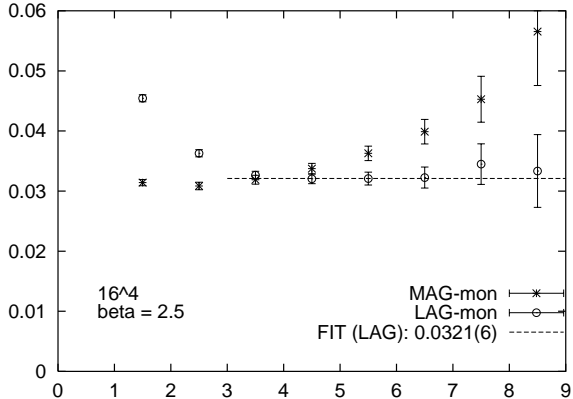}
}
\vspace*{-5mm}
\caption{
Monopole dominance:
Diagonal monopole Creutz ratios $\chi(L+\half,L+\half)$ against $L+\half$,
in MAG and LAG.
}
\vspace*{-2mm}
\label{fig2}
\end{figure}

Fig.~2 shows diagonal Creutz ratios for the monopole part of the abelian
potential, for MAG and LAG. 
There is a notable difference between the two projections.
The LAG signal decays to a plateau, which allows an excellent fit,
leading to a LAG-monopole string tension of
\be
  a^2 \sigma_{\rm mon} \ = \ 0.0321(6)
 ,
\label{sigval}
\ee
in excellent agreement with the full string tension on the same lattice
\cite{hart}.

The MAG Creutz ratios, on the other hand,
show a rising tendency.
In view of the small error bars it is difficult to argue that the data are
consistent with a plateau, and no asymptotic string tension can be extracted.

A possible explanation is as follows.
The MAG algorithm is a local iterative procedure, which does well locally
but is unable to do the global optimization well.
As a result of this defect in the gauge fixing algorithm, artificial
decorrelations in abelian projected or monopole Wilson loops might show up
at some intermediate distance,
leading to an apparently smaller correlation length, hence an apparently
larger string tension.
In other words, imperfect gauge fixing leads to abelian projected and
monopole string tensions which are {\em larger\/} than the true non-abelian
string tension.
The rising tendency of the MAG Creutz ratios in Fig.~2 with distance
might reflect precisely this effect.

The Laplacian projection, on the other hand, by construction looks at
the lowest-momentum, longest-distance eigenmode of the covariant Laplacian,
so a similar artificial intermediate-distance decorrelation is expected
(and confirmed) to be absent.

\vspace*{2.5mm}

It is a pleasure to thank Ph.~de~Forcrand for stimulating discussions.


\vspace*{2mm}

\begin{thebibliography}{99}

\bibitem{mag}
A.S. Kronfeld, M.L. Laursen, G. Schierholz and U.-J. Wiese,
{\em Monopole Condensation and Color Confinement},
\plb{198}(1987) 516.

\bibitem{born}
V.G. Bornyakov, E.-M. Ilgenfritz, M.L. Laursen, V.K. Mitrjushkin,
M. M\"{u}ller-Preussker, A.J. van der Sijs and A.M. Zadorozhny,
{\em The Density of Monopoles in SU(2) Lattice Gauge Theory},
\plb{261}(1991) 116.

\bibitem{suzukiambig}
S. Hioki, S. Kitahara, Y. Matsubara, O. Miyamura, S. Ohno and T. Suzuki,
{\em Gauge-fixing Ambiguity and Monopole},
\plb{271}(1991) 201.

\bibitem{lat96}
A.J. van der Sijs,
{\em Laplacian Abelian Projection},
\nproc{53}(1997) 535 (hep-lat/9608041).

\bibitem{ykis97}
A.J. van der Sijs,
{\em Abelian Projection without Ambiguities},
preprint SCSC-TR-98-01 (ETH-Z\"urich), hep-lat/9803001,
to appear
in the Proceedings of the 1997 Yukawa International Seminar (YKIS'97),
``Non-Perturbative QCD --- Structure of the QCD Vacuum'',
Kyoto, Japan, 2--12 December 1997.

\bibitem{thesis}
A.J. van der Sijs,
{\em Monopoles and Confinement in SU(2) Gauge Theory},
Thesis, University of Amsterdam, February 1991.

\bibitem{suzyot}
T. Suzuki and I. Yotsuyanagi,
{\em A possible evidence for Abelian dominance in quark confinement},
\prd{42}(1990) 4257.

\bibitem{shibasuzuki}
H. Shiba and T. Suzuki,
{\em Monopoles and string tension in SU(2) QCD},
\plb{333}(1994) 461.

\bibitem{hart}
A. Hart, J.D. Stack and M. Teper,
{\em The string tension in the maximally Abelian gauge after smoothing},
hep-lat/9808050.

\bibitem{smitsijs}
J. Smit and A.J. van der Sijs,
{\em Monopoles and Confinement},
Nucl.\ Phys.\ B 355 (1991) 603.

\end{thebibliography}
\end{document}